\documentclass[10pt]{article}
\usepackage[top=30mm,bottom=30mm,left=30mm,right=30mm]{geometry}
\usepackage{amsmath,ulem}
\usepackage{eucal}
\usepackage[usenames,dvipsnames]{color}
\usepackage{graphicx}
\usepackage{caption}
\usepackage{enumerate}
\usepackage{subcaption, placeins, blindtext}
\usepackage{amssymb,amsfonts}
\usepackage{bm}
\usepackage{bbm}
\usepackage{amssymb}
\usepackage{physics}
\usepackage{amsthm}
\usepackage{graphicx}
\graphicspath{ {images/} }

\usepackage{color}

\usepackage{mathtools}

\usepackage{algorithm}
\usepackage[noend]{algpseudocode}
\usepackage{mathtools}

\usepackage{graphicx}
\usepackage{dcolumn}

\makeatletter
\def\BState{\State\hskip-\ALG@thistlm}
\makeatother

\theoremstyle{definition}

\usepackage[utf8]{inputenc}
\usepackage[english]{babel}

\usepackage[titletoc,title]{appendix}

\title{Report: A Shot Noise Neuron Model}
\author{Zihao Xu}

\date{June 15, 2017}

\begin{document}

\maketitle
\section{Introduction}
In this paper, we propose a shot noise-based leaky integrated and firing neuron model and provide a detailed analysis on the performance of this model compared to the traditional diffusion approximated model. \\
In theoretical neuroscience, there are three general neuron models in the field. 
\begin{enumerate}
    \item Compartmental neuron model is a conductance-based model, in which it views the biological neurons as a large circuit. The problem of this model comes from its structure complexity and the number of its free parameters. 
    \item Leaky integrated and firing model is a more flexible model due to the special design called threshold-resetting, in which the voltage of the neuron is reset after reaching the threshold. This model is proposed as an alternative to the compartment model to provide a more biologically realistic model that can capture spike timing behaviors that are observed in experiments. In addition to that, it is more computationally efficient since it has much less free parameters. 
    \item Firing rate-based neuron model uses the firing rate of the neuron as the coupling terms in the coupled neuronal network models. Models of this kind can be easily analyzed in the network level. Theoretical analysis for this model proves the existence of phase transition and shows that the computational capacity is boosted in the chaotic regime.
\end{enumerate}
The paper is structured as the following. In section 2, some necessary background knowledge on our model is introduced. In section 3, the two modeling approaches, shot noise and diffusion approximation, are compared analytically in both steady states and perturbed states. In section 4, we discuss the advantages and drawbacks of this model and proposes future works. In the final section, a preliminary computational experiment is conducted. 
\section{Background}
In this paper, we focus on comparing two approaches, shot noise process and diffusion process, in modeling the dynamics of a neuron. Let's have a brief overview on both subjects.\\
Diffusion process is a continuous-time Markov process with continuous sample path. Due to its mathematical tractability, diffusion approximation has been applied to various processes, in which they're not originally tractable. However, this approximation is far from optimal. In later sections, we will that this approximation smooths out many dynamics.\\
Shot noise process is defined as the process that the state variables jump by positive or negative amounts irregularly. Rigorously, let's define $T_1<T_2<...$ be a sequence of arrival times of shots and $C(t), t \in \mathbb{R}_{\geq 0}$, be the noise. Then the shot-noise process $S_t$ is given by, for $t \geq 0$,
\begin{align}
S_t = \sum_{i}C(T_i)\delta(t - T_i)
\end{align}
Note that this shot noise process doesn't need to be Markovian, whereas the diffusion process is built upon the Markov assumption. As we will show later, the outcomes of this difference are significant and shot noise process is generally much better than the diffusion approximation.
\subsection{Neuron Model}
Traditionally, the biophysical neuron model is described by tons of nonlinear ion dynamics in a natural way. Here, we use a simplified model called integrated-and-firing model. Namely, the model only captures the essential behaviors of integration and firing processes of neurons by modeling the leaky membrane potential effects and the arrivals of excitatory/inhibitory spikes. The dynamics is modeled by,
\begin{align}
\dv{v}{t} = -\frac{v}{\tau} + \sum_{e} a_{ek}\delta(t - t_{ek}) + \sum_{i}a_{ik}\delta(t - t_{ik})
\end{align}
where $v$ is the voltage and $\tau$ is the time constant. This model describes the process that a point neuron receives multiple additive excitatory($a_{ek} > 0$) and inhibitory($a_{ik} < 0$) synapses at some Poisson distributed arrival times($t_{ek}, t_{ik}$). An action potential or spike happens when the voltage reaches the threshold $v_{th}$ and it is immediately reset to $v_{re}$. In the next section, we will look at the steady-state and the response properties of this shot-noise model.\\
\subsection{Neuron Model in Biological Settings}
Let's look at the probability distribution of the phase(voltage) of one single neuron in this population. With the continuity equation, we have the following,
\begin{align}
\pdv{P}{t} + \pdv{J}{v} = r(t)(\delta(v - v_{re}) - \delta(v - v_{th}))
\end{align}
where $r(t) \geq 0$ is the instantaneous firing rate of the population. In this equation, the source term, in which the neuronal voltage is generated, appears only when the neuronal voltage is reset to $v_{re}$ and the sink appears only when the voltage reaches the threshold. The flux can be modeled in terms of three components, convection, excitatory synaptic drives and inhibitory synaptic drives. Namely,
\begin{align}
J = \frac{-vP}{\tau} + J_e + J_i
\end{align}
where the excitatory flux $J_e > 0$ and the inhibitory flux $J_i < 0$.\\
In experimental neuroscience, the interspike interval, which is the distribution of the intervals between two consecutive spikes, is shown to be Poisson distributed or log-normal distributed. Therefore, the presynaptic drives are generally treated as Poisson spikes or log-normal spikes. Biologically, the presynaptic drives are transmitted to postsynaptic sites of the targeted neuron. In the targeted neuron side, the excitatory/inhibitory presynaptic drives will result in the excitatory/inhibitory postsynaptic potential(EPSP/IPSP) and these EPSP/IPSP are shown to be exponentially decayed experimentally. Therefore, the fluxes for a single neuron arising from the Poisson drives can be intuitively modeled as the integrals over the postsynaptic voltage density and the synaptic-amplitude distributions. Therefore, the fluxes are given by,
\begin{align}
J_e(v,t) = R_e(t)\int_{-\infty}^v dw P(w,t)\int_{v-w}^{\infty}da A_e(a)\\
J_i(v,t) = R_i(t)\int_{-\infty}^v dw P(w,t)\int_{v-w}^{\infty}da A_i(a)
\end{align}
where $R_e(t)$, $R_i(t)$ are excitatory/inhibitory firing rate and $A_e$ and $A_i$ are the synaptic-amplitude distributions. Explicitly, $A_e(a) = e^{-a/a_e}/a_e$ and $A_i(a) = -e^{-a/a_i}/a_i$. Note that in our neuronal model (2), the resting potential is $0$mV, so the mean postsynaptic potentials are $a_e > 0$ and $a_i < 0$.
\begin{align}
J_e(v, t) = R_e(t)\int_{-\infty}^v dw P(w, t) e^{-(v - w)/a_e}
\end{align}
Note that it is the Fredholm integral equation of the first kind, convolving with an exponential kernel. Therefore, it can be rewritten into a linear differential form.
\begin{align}
\pdv{J_e}{v} &= -\frac{R_e}{a_e}\int_{-\infty}^v dw (P(w, t) e^{-(v - w)/a_e}) + R_eP
\end{align}
For the excitatory case, we need to treat our neuronal model carefully. Remember that the excitatory spike is induced when the voltage reaches the threshold. At that point, the excitatory flux is equal to the firing rate and the inhibitory flux is equal to zero. That is $J_e(v_{th}) = J(v_{th}) = r$ and $J_i(v_{th})$. Put these quantities into equation (4). We see that in this case $P = 0$, which says that the integrals in (5) doesn't capture this feature. Therefore, the excitatory flux needs another correction term, $r\delta(v - v_{th})$. So,
\begin{align}
\pdv{J_e}{v}  &= -\frac{J_e}{a_e} + R_eP - r(\delta(v - v_{th}))\\
\pdv{J_i}{v} &=-\frac{J_i}{a_i} + R_iP
\end{align}
Since equations(3,4,9,10) are all linear in fluxes an probability density, we can apply Laplace transforms to analytically solve for the firing rate of the model.
\section{Shot noise process v.s. Diffusion Approximation}
\subsection{Steady State}
\subsubsection{Sub-threshold Moment}
The goal of this paper is to quantitatively compare the difference between the shot-noise process and the diffusion approximation. For the diffusion approximation, the mean and the variance of the pdf are required quantities in the dynamics, so we start with computing the mean and the variance of the pdf in the tractable case, the sub-threshold limit. Let's take the sub-threshold limit,$v_{th} \to \infty$). With this limit, we can easily see that the right hand side of the continuity equation becomes zero($r = 0$) and the voltage is conserved. Therefore, the flux becomes zero $J = 0$ as well. Note that the bilateral Laplace transformation of the probability distribution function is the moment generating function for the random variables drawn from that pdf. Now we are ready to use bilateral Laplace transform, $\tilde f(s) = \int_{-\infty}^\infty dv e^{sv}f(v)$, to the equations under this limit.
For the excitatory and inhibitory fluxes equations (9),(10) and continuity equations,
\begin{align}
\tilde J_0 &= 0\\
-s\tilde J_{e0} &= -\frac{\tilde J_{e0}}{a_e} + R_{e0}Z_0\\
-s\tilde J_{i0} &= -\frac{\tilde J_{i0}}{a_i} + R_{i0}Z_0
\end{align}
where subscript 0 stands for the quantity in the steady state.\\
By plugging these quantities into (4), we have,
\begin{align}
\tilde J_0(s) &= (\dv{Z_0}{s})/\tau+ \tilde J_{e0} + \tilde J_{i0}\\
\dv{Z_0}{s} &= (\frac{a_eR_{e0}}{1 - a_es} + \frac{a_iR_{i0}}{1 - a_is})Z_0 \tau
\end{align}
By integrating the above equation with the unity condition that $\tilde P_0(0) = 1$, we obtain,
\begin{align}
Z_0 = \frac{1}{(1 - a_es)^{\tau R_{e0}}(1 - a_is)^{\tau R_{i0}}}
\end{align}
By taking $W_0 = \log Z_0$ and computing the first and second moments, $\dv{W_0}{s}|_{s=0}$ and $\frac{d^2 W_0}{ds^2}|_{s=0}$, we reach the mean $\mu_0$ and variance $\sigma_0^2$ for the subthreshold limit,
\begin{align}
\mu_0 &= a_e\tau R_{e0} + a_i \tau R_{i0}\\
\sigma_0^2 &= a_e^2\tau R_{e0} + a_i^2\tau R_{i0}
\end{align}
\subsubsection{Shot Noise Solution}
We apply bilateral Laplace transforms to the continuity equation and its related equations in the steady state $\pdv{P}{t} = 0$. For the continuity equation (3), we have,
\begin{align}
\int_{-\infty}^\infty dv e^{sv}\pdv{J}{v} &= r_0 \int_{-\infty}^\infty dv e^{sv}(\delta(v - v_{re}) - \delta(v - v_{th}))\\
s\tilde J_0(s) &= r_0 (e^{sv_{re}} - e^{sv_{th}})\\
\tilde J_0(s) &= \frac{r_0}{s} (e^{sv_{re}} - e^{sv_{th}})
\end{align}
For the excitatory and inhibitory fluxes equations (9),(10),
\begin{align}
-s\tilde J_{e0} &= -\frac{\tilde J_{e0}}{a_e} + R_{e0}\tilde P_0 - r_0 e^{sv_{th}}\\
-s\tilde J_{i0} &= -\frac{\tilde J_{i0}}{a_i} + R_{i0}\tilde P_0
\end{align}
Combine the results above to the Laplace transformed flux equations (4),
\begin{align}
\tilde J_0(s) &= (\dv{\tilde P_0}{s})/\tau+ \tilde J_{e0} + \tilde J_{i0}\\
\frac{r_0}{s} (e^{sv_{re}} - e^{sv_{th}}) &= (\dv{\tilde P_0}{s})/\tau - a_e(R_{e0}\tilde P_0 - r_0 e^{sv_{th}})/(1 - sa_e) - a_iR_{i0}\tilde P_0/(1 - sa_i)\\
\dv{\tilde P_0}{s} &= (a_eR_{e0}/(1 - sa_e) + a_iR_{i0}/(1 - sa_i))\tau\tilde P_0 - \frac{r_0\tau}{s} (\frac{e^{sv_{th}} }{1 - sa_e}-e^{sv_{re}})
\end{align}
From the results in (15), it's clear that the bracket term in the RHS of (26) can be rewritten as $\frac{1}{Z_0}\dv{Z_0}{s}$, then
\begin{align}
\dv{\tilde P_0}{s} &= \frac{1}{Z_0}\dv{Z_0}{s}\tilde P_0 - \frac{r_0\tau}{s} (\frac{e^{sv_{th}} }{1 - sa_e}-e^{sv_{re}})
\end{align}
Multiplying $e^{log(\frac{1}{Z_0})}$ to BHS and integrating between $s$ and $1/a_e$, we get,
\begin{align}
\tilde P_0 = \tau r_0 \int_s^{1/a_e} \frac{dc}{c} \frac{Z_0(s)}{Z_0(c)} (\frac{e^{cv_{th}} }{1 - ca_e}-e^{cv_{re}})
\end{align}
With the unity condition trick $\tilde P_0(0) = 1 = Z_0(0)$ at $s = 0$,
\begin{align}
\frac{1}{\tau r_0}=\int_0^{1/a_e} \frac{dc}{c} \frac{1}{Z_0(c)} (\frac{e^{cv_{th}} }{1 - ca_e}-e^{cv_{re}})
\end{align}
In the diffusion limit of small synaptic amplitudes($a_e \to 0$) and reasonably high firing rate, we can easily derived the form by approximate the moment generating function to the order of the variance.
\begin{align}
\frac{1}{\tau r_0} \approx \int_0^{\infty} \frac{dy}{y} e^{-y^2/2} (e^{yy_{th}} - e^{yy_{re}})
\end{align}
where $y_{th} = (v_{th} - \mu_0)/\sigma_0$ and $y_{re} = (v_{re} - \mu_0)/\sigma_0$.\\
\subsubsection{Diffusion Approximation}
The key idea of the diffusion approximation is to use the mean and the variance of the probability distribution function to capture the leading dynamics. In the diffusion approximation, the voltage dynamics and the flux equations are described by,
\begin{align}
\tau\dv{v}{t} &= \mu - v + \sigma \sqrt{2\tau} \eta(t)\\
\tau J &= (\mu - v) P - \sigma^2 \dv{P}{v}
\end{align}
where $\eta$ is a zero mean and unit variance white noise. By doing the bilateral Laplace transform on (29) in the steady state and plugging it into (21), we get,
\begin{align}
\tau \tilde J_0 &= \mu \tilde P_0 + \dv{\tilde P_0}{s} + \sigma^2 s \tilde P_0\\
\tilde J_0(s) &= \frac{r_0}{s} (e^{sv_{re}} - e^{sv_{th}})\\
\dv{\tilde P_0}{s} &= (\mu_0 + \sigma_0^2 s)\tilde P_0 +  \frac{r_0\tau}{s} (e^{sv_{re}} - e^{sv_{th}})
\end{align}
By solving this first order linear ODE, we get,
\begin{align}
d\tilde P_0 e^{-\mu_0s - \frac{1}{2}\sigma_0^2 s^2} = ds e^{-\mu_0s - \frac{1}{2}\sigma_0^2 s^2}\frac{r_0\tau}{s} (e^{sv_{re}} - e^{sv_{th}})\\
\tilde P_0 = r_0\tau e^{\mu_0s + \frac{1}{2}\sigma_0^2 s^2} \int_s^{\infty} dx e^{-\mu_0x - \frac{1}{2}\sigma_0^2 x^2}\frac{1}{x} (e^{xv_{th}} - e^{xv_{re}})
\end{align}
Take $s = 0$, then $\tilde P_0(0) = 1$, and the steady state firing rate $r_0$ follows,
\begin{align}
\frac{1}{r_0\tau} &= \int_0^{\infty} dx e^{-\mu_0x - \frac{1}{2}\sigma_0^2 x^2}\frac{1}{x} (e^{xv_{th}} - e^{xv_{re}})\\
&= \int_0^{\infty} \frac{dy}{y} e^{-y^2/2} (e^{yy_{th}} - e^{yy_{re}})
\end{align}
where $y_{th}$ and $y_{re}$ are defined in the previous section.\\
The equation (30) and (39) coincidentally match up. The diffusion approximation can capture the dynamics of the shot-noise process only up to the order of the variance. It is due to the fact that the dynamics used in the diffusion approximation only uses the information of the mean and the variance of the voltage pdf.
\subsection{Neuronal Firing Rate Responses to Modulated Presynaptic Rates}
The steady state firing rates, $r_0$, for both models share some similar structures. However, the differences are not very significant. We conjecture that this is due to the reason that the firing rate at the steady state smooths out almost all sharp waves in order to be in the steady state. Therefore, an obvious second step is to study firing rates in the non-steady-state case. \\
The easiest and the most straightforward way to study the non-steady-state behavior is to perturb the neuron in the steady state. This gives us several more options to play with, perturbing the total fluxes. With the above reasoning at hand, we choose to modulate the presynaptic rates. Namely, adding a periodic waves to the steady-state presynaptic rates. This provides us a way to look at the neuronal firing rate responses at different levels around the original steady state rate.
\subsubsection{Shot Noise Solution}
Remember that the firing responses in the neuron are determined by the firing rates at the presynaptic sites and the responses of the EPSP/IPSP, equation (5),(6). The responses of the EPSP/IPSP curves cannot be changed. Otherwise, the model loses its biological meaning. So the only option left here is the perturbation on the presynaptic rate in the steady state. Let's modulate either excitatory/inhibitory presynaptic rates and the resulted neuronal rate responses are written as $r_e$ or $r_i$. So the modulated presynaptic rates become $R_{e} = R_{e0} + R_{e1}e^{iwt}$ or $R_{i} = R_{i0} + R_{i1}e^{iwt}$. This modulation naturally gives the modulation on the pdf of voltage up to the first order, $P = P_0 + P_1 e^{iwt}$. \\
By redoing the analysis above, bilateral Laplace transformations, we reach at,
\begin{align}
\dv{\tilde P_1}{s} &= (\frac{a_eR_{e0}}{1 - sa_e} + \frac{a_iR_{i0}}{1 - sa_i}-\frac{iw }{s})\tau\tilde P_1 +  \frac{a_kR_{k1}}{1 - sa_k}\tau\tilde P_0 - \frac{r_k\tau}{s} (\frac{e^{sv_{th}} }{1 - sa_e}-e^{sv_{re}})
\end{align}
where $r_k$'s are modulated firing rates that are of our interests and Einstein's summation convention is used here.\\
Take $Z_1 = \frac{Z_0}{s^{iw\tau}}$. Then we can simplify the equation above in the same manner that we did previously.
\begin{align}
\dv{\tilde P_1}{s} &= (\frac{1}{Z_0}\dv{Z_0}{s}-\frac{iw \tau}{s})\tilde P_1 +  \frac{a_kR_{k1}}{1 - sa_k}\tau\tilde P_0 - \frac{r_k\tau}{s} (\frac{e^{sv_{th}} }{1 - sa_e}-e^{sv_{re}})\\
\dv{\tilde P_1}{s} &= (\frac{1}{Z_1}\dv{Z_1}{s} + \frac{iw\tau Z_1 s^{iw\tau - 1}}{Z_1 s^{iw\tau}}-\frac{iw \tau}{s})\tilde P_1 +  \frac{a_kR_{k1}}{1 - sa_k}\tau\tilde P_0 - \frac{r_k\tau}{s} (\frac{e^{sv_{th}} }{1 - sa_e}-e^{sv_{re}})\\
\frac{1}{Z_1}\dv{}{s}\frac{\tilde P_1}{Z_1} &= \frac{a_kR_{k1}}{1 - sa_k}\tau\tilde P_0 - \frac{r_k\tau}{s} (\frac{e^{sv_{th}} }{1 - sa_e}-e^{sv_{re}}) 
\end{align}
Again, this is the first order linear ODE. Integrate over $s$ and $1/a_e$ in the limit that $s \to 0$. Under this limit, $\tilde P(0) = \tilde P_0(0) = 1$ and $\tilde P_1 = 0$. Then we get,
\begin{align}
r_k &= R_{k1}\frac{\int_0^{1/a_e} \frac{ds a_k}{1 - a_k s}\frac{1}{Z_1} \tilde P_0}{\int_0^{1/a_e} \frac{ds}{s} \frac{1}{Z_1} (\frac{e^{sv_{th}} }{1 - sa_e}-e^{sv_{re}})}\\
&= R_{k1}\frac{\int_0^{1/a_e} \frac{ds a_k}{1 - a_k s}\frac{1}{Z_1} \tau r_0 \int_s^{1/a_e} \frac{dc}{c} \frac{Z_0(s)}{Z_0(c)} (\frac{e^{cv_{th}} }{1 - ca_e}-e^{cv_{re}})}{\int_0^{1/a_e} \frac{ds}{s} \frac{1}{Z_1} (\frac{e^{sv_{th}} }{1 - sa_e}-e^{sv_{re}})}\\
&= R_{k1}\tau r_0 \frac{\int_0^{1/a_e} \frac{ds a_k}{1 - a_k s}s^{iw \tau} \int_s^{1/a_e} \frac{dc}{c} \frac{1}{Z_0(c)} (\frac{e^{cv_{th}} }{1 - ca_e}-e^{cv_{re}})}{\int_0^{1/a_e} \frac{ds}{s} \frac{s^{iw\tau}}{Z_0} (\frac{e^{sv_{th}} }{1 - sa_e}-e^{sv_{re}})}\\
&= R_{k1}\tau r_0 \frac{\int_0^{1/a_e} \frac{dc}{c} \frac{1}{Z_0(c)} (\frac{e^{cv_{th}} }{1 - ca_e}-e^{cv_{re}})\int_0^{c} \frac{ds a_k}{1 - a_k s}s^{iw \tau}}{\int_0^{1/a_e} \frac{ds}{s} \frac{s^{iw\tau}}{Z_0} (\frac{e^{sv_{th}} }{1 - sa_e}-e^{sv_{re}})}
\end{align}
Low frequency limit only cares about the steady-state firing rate as we expected, so let's move onto the high frequency limit scenario. In order to do calculations in the high frequency limit, we need to use some algebraic tricks, taking the substitution $s = e^{-x}/a_e$, rotating in the complex plane $x \to q/iw\tau$ and expanding in powers of inverse frequency. With some tedious algebra, we will obtain the a series of gamma function and the results follow by approximating the series up to order $\mathcal{O}(1/w^2)$.\\
For the excitation modulations, the firing rate response is,
\begin{align}
r_e \approx r_0 \frac{R_{e1}}{R_{e0}}(1 - \frac{R_{e0}\tau}{iw\tau}(\frac{c_1}{R_{e0}\tau + 1}+ \frac{1}{2}))
\end{align}
where $c_1 = \frac{1}{2} - \frac{v_{th}}{a_e} - e^{-(v_{th} - v_{re})/a_e} - \frac{R_{e0}\tau}{2} + \frac{a_i R_{i0} \tau}{a_e - a_i}$.\\
For the inhibition modulations,
\begin{align}
r_i \approx r_0 \frac{R_{i1}\tau}{iw\tau}\frac{a_i}{a_e - a_i}(1 - \frac{R_{e0}\tau + 1}{iw\tau}\frac{a_e}{a_e - a_i})
\end{align}
\subsubsection{Diffusion Approximation}
In the diffusion approximation model, we can modulate either on the mean $\mu = \mu_0 + \mu_1 e^{iwt}$ or the variance $\sigma^2 = \sigma_0^2 + \sigma_1^2 e^{iwt}$. Then the modulated flux becomes,
\begin{align}
\tau J_1 &= P_1 (\mu_0 - v) - \sigma_0^2 \dv{P_1}{v} + P_0 \mu_1\\
\text{or } \tau J_1 &= P_1(\mu_0 - v) - \sigma_0^2 \dv{P_1}{v} - \sigma_1^2 \dv{P_0}{v}
\end{align}
Redo the same algebraic tricks as we did on the shot-noise case. We obtain the  integral form.
\begin{align}
B_m = \int_0^{\infty} \frac{dy}{y}y^{m + iw\tau}e^{-y^2/2}(e^{yy_{th}} - e^{yy_{re}})    
\end{align}
Therefore, the rates for mean and variance can be obtained as the following,
\begin{align}
r_\mu = \frac{r_0}{1 + iw\tau}\frac{\mu_1}{\sigma_0}\frac{B_1}{B_0} \text{ and } r_\sigma = \frac{r_0}{2 + iw\tau}\frac{\sigma_1^2}{\sigma_0^2}\frac{B_2}{B_0}
\end{align}
Note that either excitatory modulation or inhibitory modulation could affect both the mean and the variance (17), (18). Therefore, a modulation in the shot noise model naturally gives a simultaneous mean and variance modulation in the diffusion approximation model. By taking the modulation effects up to the first order, we can approximate the rates by $r_k = r_\mu + r_\sigma$. Then,
\begin{align}
r_k \approx r_0(\frac{\sigma_1^2}{\sigma_0^2} + \frac{1}{\sqrt{iw\tau}}(\frac{\mu_1}{\sigma_0} + \frac{\sigma_1^2}{\sigma_0^2}\frac{v_{th}-\mu_0}{\sigma_0}))
\end{align}
where $\mu_1 = R_{k1}\tau a_k$ and $\sigma_1^2 = R_{k1}\tau a_k^2$.\\
The modulation effects make the distinctions between the shot-noise process and the diffusion approximation more obvious. The diffusion approximation doesn't distinguish the inhibitory rates and the excitatory rates. This problem arises from the limitation of the diffusion approximation approach, in which it uses only the information of the mean and the variance of the pdf. This treatment makes the dynamics unaware of the differences of the excitations and inhibitions. In the shot-noise solution, inhibitory modulations and excitatory modulations can produce different firing rates. This difference in the shot noise model naturally comes from our design of the excitatory and inhibitory fluxes dynamics. Another important point here is that shot noise rates give a scaling rule $\sim \frac{1}{iw\tau}$, whereas the diffusion approximated rates only has $\sim \frac{1}{\sqrt{iw\tau}}$. In both senses, the shot noise dynamics provides wealthier dynamics than the diffusion approximated dynamics.
\section{Discussion}
From the analysis we did above, even though the analysis is simple, the setup for the shot-noise model requires some extra attentions to the structure and threshold response of the biological neurons. One key difficulty in the shot-noise model is that the model is built upon recent results in the experimental neuroscience, the rate responses properties of neurons in the both presynaptic site and the postsynaptic site. This was not known at the time when the diffusion approximation approach was applied. Other than this, the shot-noise model has several strong advantages over the diffusion dynamics.
\begin{enumerate}
    \item The dynamics in the shot noise process is wealthier. The dynamics described by the diffusion approximation approach is essentially the low order terms in the dynamics of the shot noise process.
    \item The model doesn't care about the pdf of voltage. The major problems of the diffusion approximation approach comes from the fact that it requires the analytic solutions of the mean and the variance of the voltage pdf, whereas, in most of the time, these properties are not analytically tractable.
    \item The computational cost of the model is more efficient. Shot-noise process uses either on or off decision at each step instead of computing the whole dynamics. This greatly reduces the computational complexity of the simulations and allows computational neuroscientists to build larger and more complex networks.
\end{enumerate}
In the paper, the new shot noise model is proposed and proved to be more biological than the diffusion model. This opens up a new way to look at the neuronal network based on the shot noise model. There are several possible future works in this direction. 
\begin{enumerate}
    \item Theoretically, analyze this neuron model with multiplicative shot noise. Currently, the model uses additive shot noises for simplicity and algebraic tractability. A more biologically model could use an multiplicative factor $(\epsilon - v)$ to regulate the effects of the shot noise, where $\epsilon$ is the resting potential of the neuron.
    \item Computationally, build up a network model based on this neuron model. Due to its simplicity, a network model with different connection rules(e.g. type-to-type connections and nearest neighbor connections) can be further tested and studied. 
\end{enumerate}
\section{Computational Experiments}
In experimental neurobiology, an emergent behavior, neuronal avalanches, has been reported in cortical slices of rats somatosensory cortex in vitro, in which the distribution of the number of active neurons at one time step follows power law. In this paper, we take the computational advantages of the shot noise model introduced above to further test its performance in the network level.\\
\subsection{Method}
Avalanche process is first observed in the sandpile model, which uses the 2-D nearest neighbor interaction. For our network architecture, we consider the 2-D nearest neighbor connection as well. Other than this, we further simplify the model in the following two ways, static network and weighted connections. Neurobiologically, the number and the weights of synaptic contacts between two connected neurons vary dynamically due to connectivity learning rules and developmental rules. Here we reduce the model to be static and only consider a weighted connections(excitatory or inhibitory) between two neurons by randomly sampling 100 Gaussian random variables with mean sampled from the type of the connection(excitatory or inhibitory) and summing them up. Note that these weights determine the intensity of the shot noise.\\
In our experiments, four different connection conditions have been studied, with/without recurrent connections and excitatory-dominant or inhibitory-dominant connections. Extremal conditions are not always desired, excitatory-only/inhibitory-only connections. It's not hard to see that excitatory-only connection is nothing more than a diffusion process and inhibitory-only connection kills the whole network at the very beginning. Therefore, excitatory-dominant or inhibitory-dominant(E/I) connections conditions are studied here. Furthermore, we introduce recurrent connection mechanisms to nearly half of the neurons in the network. Recurrent connection mechanism serves as the self-regularization(SR) process, in which the excitatory recurrent connection is applied if the original network is inhibitory-dominant and vice versa. The stimulus are given after the network reaches its steady state by injecting currents to the first neuron(indexed as (0,0) in the grid) in a very short period.\\
\subsection{Analysis and Results}
\begin{figure}[h]
    \centering
    \begin{subfigure}{.5\textwidth}
    \centering
     \includegraphics[width=1.1\linewidth]{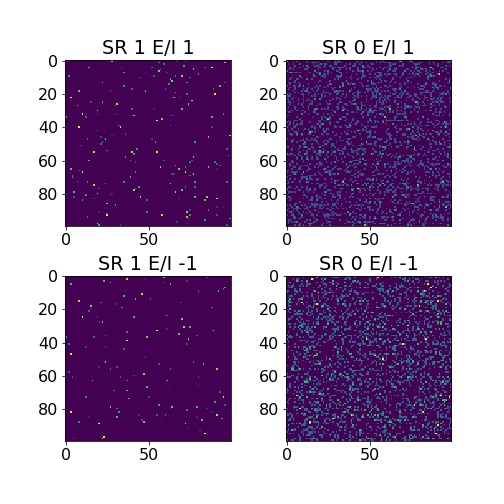}
     \caption{Firing Rate Grid Plot}
     \label{fig:sub1}
    \end{subfigure}%
    \begin{subfigure}{.5\textwidth}
     \centering
    \includegraphics[width=1.1\linewidth]{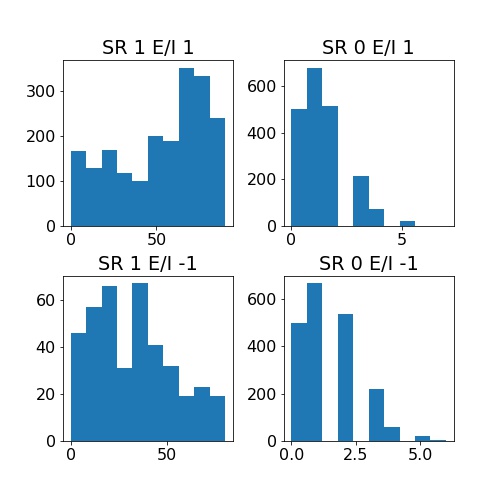}
    \caption{Histogram of the number of avalanches at one time step(x-axis)}
    \label{fig:sub2}
    \end{subfigure}
    \caption{The network model uses 100 by 100 neurons in a 2D grid. SR stands for recurrent connection, where 1/0 is on/off, and 1/-1 (in E/I) means excitatory-dominant or inhibitory dominant. The color bar for (a) goes from violet[0] to yellow[1]. Each subplot in (a) is rescaled to 1.}
    \label{fig:mesh1}
\end{figure}
\begin{figure}[h]
    \centering
    \includegraphics[width=1.1\linewidth]{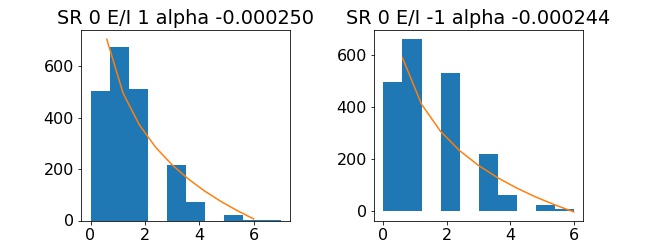}
    \caption{}
    \label{fig:mesh1}
\end{figure}
The experiment results are shown in Figure 1. In Figure 1 (a), we show the 2D grid plot of the firing rate for each neuron. This plot shows that the spikes in the network are generally sparse and this agrees with both in-vitro and in-vivo experiments. However, the firing rate patterns of excitatory-dominant condition and inhibitory-donimant condition are largely different. In the excitatory-dominant condition, we can see that there are some dominant sources that constantly generate spikes, whereas in the inhibitory-dominant case spiking events are rare. In figure 1 (b), the histogram of avalanches events is presented. From the figure, the self-regularization process(recurrent connection) kills the power law-like curve. After fitting the shape under the non-recurrent connection conditions to the curve, Figure 2, we see that the exponents are surprisingly low($\ll 1$) compared to the exponents obtained in the sandpile model($\sim 1$). These difficulties observed in our experiments might come from the following factors.
\begin{enumerate}
    \item Too many free parameters in the model. The correct choice of free parameters are not known at this point due to the scale of the network. We cannot simply use the parameters observed in experiments, since the scale of the network and the accuracy of the neuronal model are different.
    \item The non-selective(nearest neighbor) connection rule. In vitro experiments have shown that cell-type specific connection rule is widely observed in brain and the activity of the neuronal networks is sparse. Possible improvements on this direction are using a more sparse network, in which the network is a low-degree graph, or making several different shot noise neurons with different firing responses.
    \item Static Network. One can further add dynamic features into this model, synaptic weight learning rule and synapse formation rule.
\end{enumerate}

M. J.E. Richardson and R. Swarbrick
Phys Rev Lett, 105, 178102(2010) 
\end{document}